\documentclass[%
 aip,
 amsmath,amssymb,
 reprint,%
]{revtex4-1}
\usepackage{graphicx}% Include figure files
\usepackage{dcolumn}% Align table columns on decimal point
\usepackage{bm}% bold math
\usepackage[utf8]{inputenc}
\usepackage[T1]{fontenc}
\usepackage{mathptmx}
\usepackage{etoolbox}

\makeatletter
\def\@email#1#2{%
 \endgroup
 \patchcmd{\titleblock@produce}
  {\frontmatter@RRAPformat}
  {\frontmatter@RRAPformat{\produce@RRAP{*#1\href{mailto:#2}{#2}}}\frontmatter@RRAPformat}
  {}{}
}%
\makeatother
\begin{document}

\preprint{AIP/123-QED}

\title{Temperature-Dependent Emission Polarization in GaN Defect-Based Quantum Emitters}
% Force line breaks with \\
\author{Yifei Geng} % Write as First name Surname
 \email{yg474@cornell.edu}
\affiliation{School of Electrical and Computer Engineering, Cornell University, Ithaca, New York 14853, USA.}

\date{\today}% It is always \today, today,

\begin{abstract}
GaN defect-based quantum emitters show significant potential for quantum information technologies, yet their intrinsic nature is not fully understood. In this work, we present results on the temperature-dependent emission polarization of GaN defect single-photon emitters integrated with solid immersion lenses. The photoluminescence (PL) remains linearly polarized over the temperature range of 10K to 300K, with a slight rotation in the polarization direction observed at intermediate temperatures. Possible mechanisms underlying this behavior are analyzed, and a roadmap for future research is outlined.
\end{abstract}

\maketitle

GaN defect-based quantum emitters have gained increasing research interest in recent years, emerging as promising single-photon sources for quantum applications\cite{berhane2018photophysics,luo2025gan,berhane2016bright,luo20241room,geng2024defect,luo2024data,luo2023room,geng2022decoherence}. GaN defect emitters exhibit several remarkable properties, including high brightness, linear polarization, strong zero-phonon line (ZPL) emission, and the ability to operate at room temperature. In addition to these advantages, GaN, a mature direct-bandgap semiconductor widely used in photonic and electronic devices, enables the seamless integration of classical and quantum optics for on-chip applications\cite{wasisto2019beyond,yi2019topological,najda2022gan}. Recently, significant efforts have been devoted to investigating GaN defect emitters, with studies reporting on their PL spectrum, optical dephasing mechanisms, optical dipole orientation, ultrafast spectral diffusion, and spin properties\cite{geng20221temperature,geng2023ultrafast,luo2024room,geng2022temperature}. However, as a quantum emitter still in its early stages, the nature of these defects remains unknown.

Understanding the nature of defects in GaN is central to the study of defect-based quantum emitters\textemdash not only because it governs all the physics of the emitter, but also because, once the nature is known, it becomes possible to intentionally create them in GaN crystals for future quantum applications. Both first-principles calculations and experimental investigations\textemdash such as measurements of the PL spectrum, lifetime, dipole orientation, spin structure, and more\textemdash can provide significant information about their nature. Notably, the polarization characteristics of the emitted light are crucial for studying defect emitters, as they provide valuable insights into their underlying origins. GaN defect emitters are known to exhibit the characteristics of a single optical dipole at room temperature, emitting linearly polarized light with an optical dipole moment that is nearly perpendicular to the c-axis of the wurtzite crystal. Considering crystal and emission symmetry, the nature of these defects is likely complex structures composed of at least two or more entities, such as impurity atoms or impurity atom–vacancy pairs. However, despite its importance in further understanding the nature of the defect, the temperature dependence of emission polarization has yet to be reported.

In this work, we investigate the temperature-dependent polarization of light emitted from GaN defect quantum emitters integrated with solid immersion lenses, over a temperature range from 10 K to 300 K. Our experimental results show that at low temperatures, the emitted light maintains a high degree of linear polarization, consistent with the single optical dipole model. However, the polarization direction does not remain aligned with that observed at room temperature; instead, it undergoes a noticeable rotation. This observation suggests that the excited state of the GaN defect emitter is not an orbital doublet, fundamentally distinguishing it from other defects such as the NV center in diamond. Potential directions for future experimental investigations are also discussed.

\begin{figure}
\includegraphics{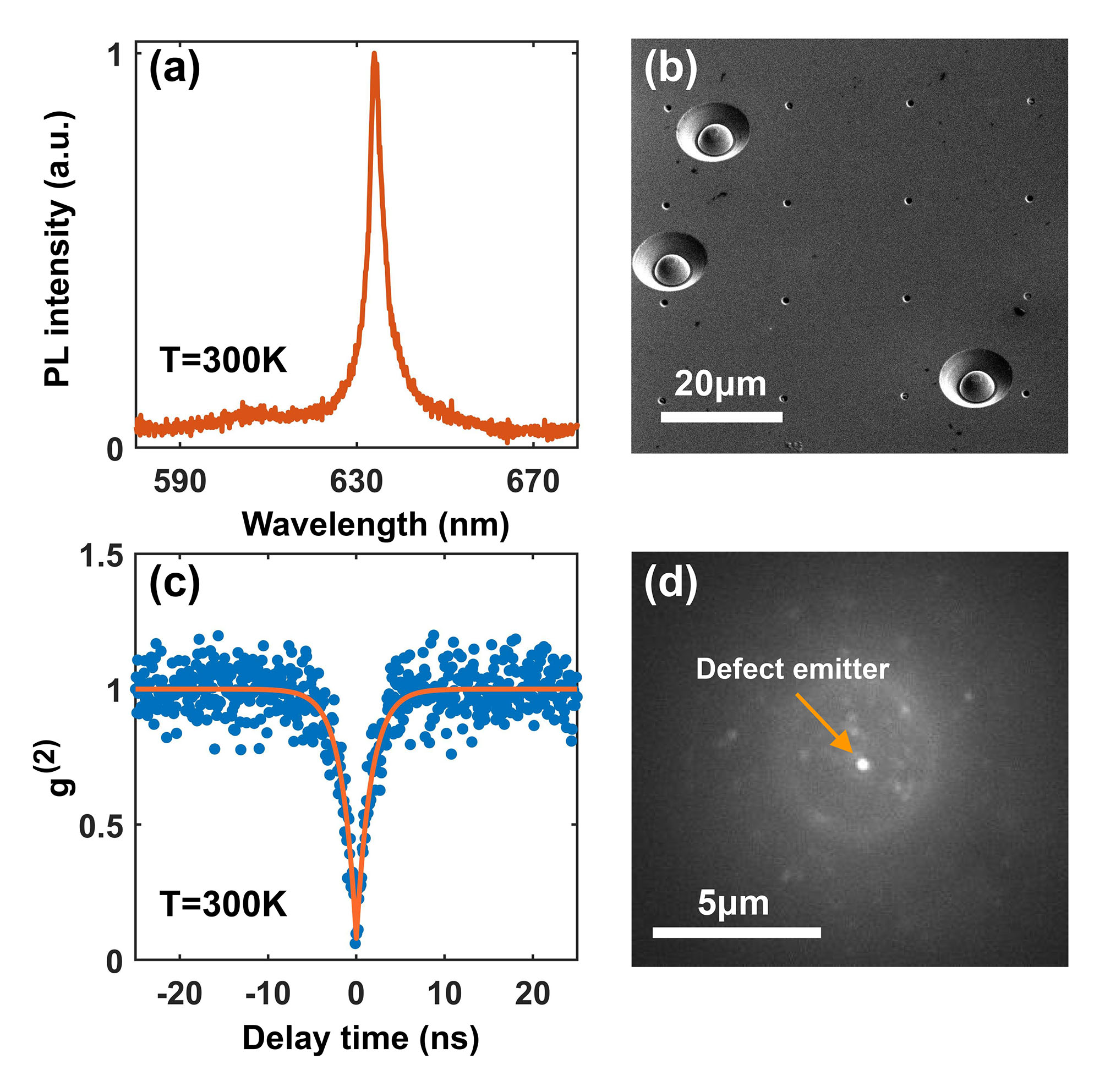}
\caption{\label{fig1} 
(a) PL spectrum of the GaN defect emitter at room temperature, with a peak emission wavelength of 634 nm. (b) Representative SEM image of fabricated solid immersion lenses (SILs) using focused ion beam (FIB) etching. The SILs are hemispherical with a diameter of 5 $\mu$m, the scale bar represents 20 $\mu$m. (c) Second-order correlation function $g^{(2)}(\tau)$ of the defect emitter measured at room temperature. (d) Spatial PL map of the defect emitter integrated within a solid immersion lens.}
\end{figure}

\begin{figure*}
\includegraphics[width=0.7\textwidth]{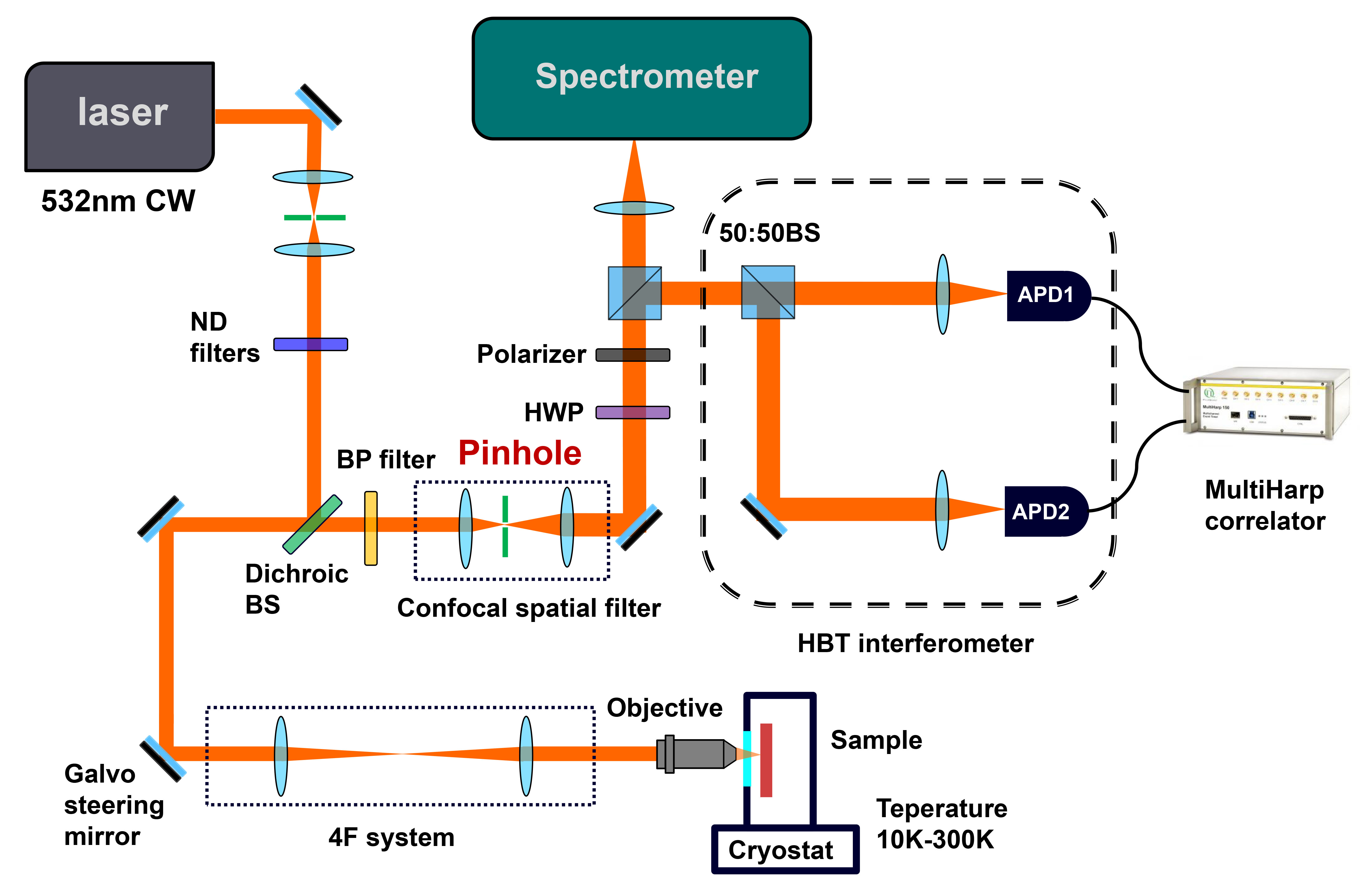}% Here is how to import EPS art
\caption{\label{fig2} Schematic layout of the custom-built confocal scanning microscope, integrated with a spectrometer and an HBT interferometer. A half-wave plate and a polarizer are placed in the collection path to analyze the polarization characteristics of the PL emission. The GaN sample is mounted inside a cryostat with temperature control ranging from 10 K to 300 K.}
\end{figure*}

\begin{figure*}[!t]
    \includegraphics[width=0.75\textwidth]{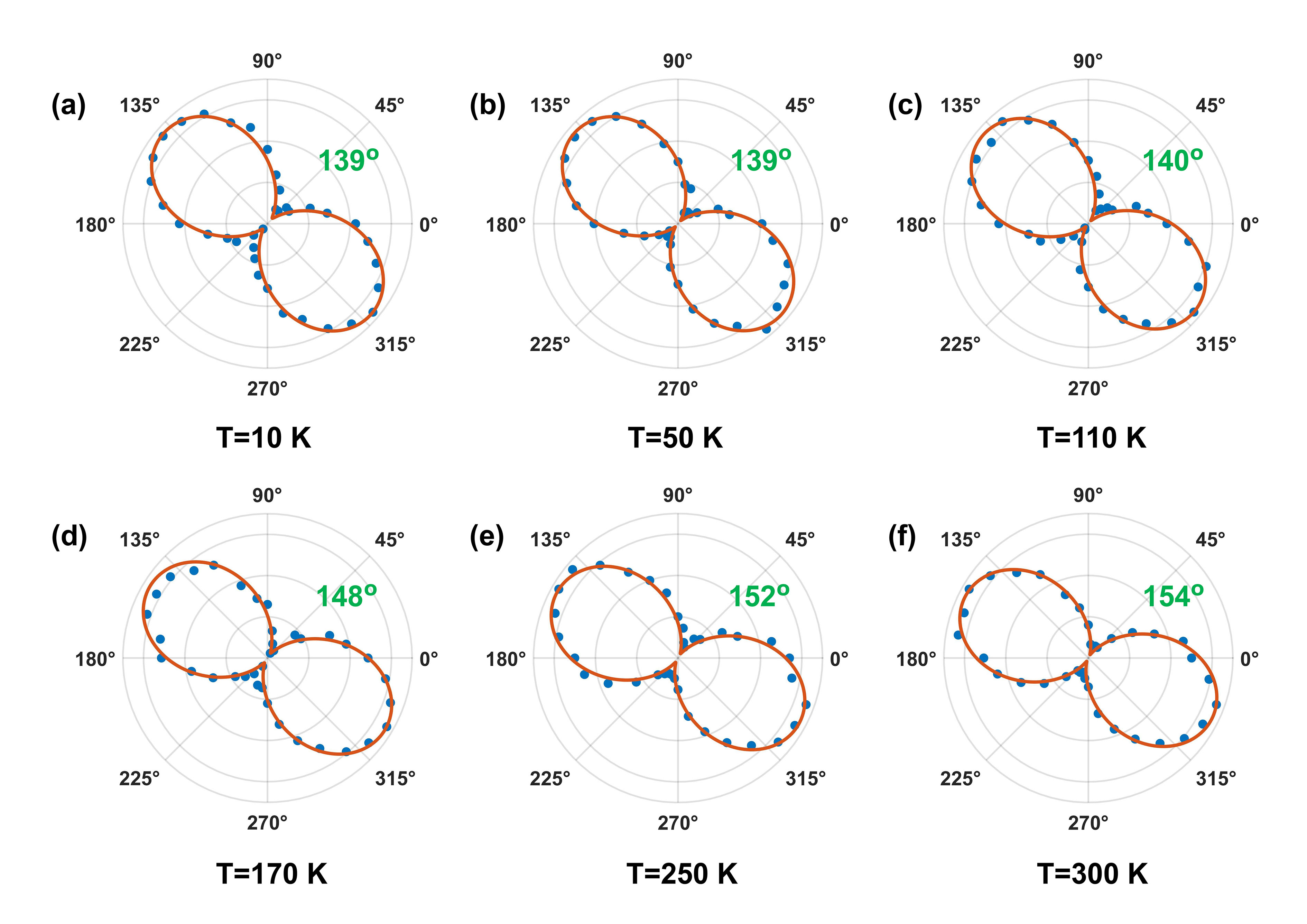}
    \centering
    \caption{Temperature-dependent emission polarization of the GaN defect emitter. Polarization directions are shown at 10 K, 50 K, 110 K, 170 K, 250 K, and 300 K.}
    \label{fig3}
    \end{figure*}

The GaN sample used in this study consists of a 4 $\mu$m-thick semi-insulating GaN layer grown by HVPE on a sapphire substrate. Figure \ref{fig1}(a) shows the PL spectrum of the GaN defect emitter under investigation at room temperature, with a peak emission wavelength of 634nm. Figure \ref{fig1}(c) presents the measured second-order correlation function $g^{(2)}(\tau)$ using a Hanbury Brown and Twiss (HBT) interferometer at room temperature. The value of $g^{(2)}(\tau=0)$ is below 0.5, confirming the single-photon emitter identity of the defect. GaN is a well-known high-refractive-index material, which is advantageous for light mode confinement in integrated photonics, but poses challenges for light extraction. The refractive index of GaN in the visible wavelength range is approximately 2.4, which causes most photons emitted from the defect emitter to undergo total internal reflection at the GaN–air interface. To enhance photon collection efficiency, a solid immersion lens (SIL) in the form of a hemisphere with a diameter of 5 $\mu$m is fabricated on top of the defect emitter using focused ion beam (FIB) etching. To prevent surface charge accumulation during the FIB process, the GaN surface is coated with a thin layer of aluminum, which can later be removed by wet etching. When the defect emitter is positioned near the center of the SIL, the collection efficiency can typically be enhanced by a factor of up to 5. Figure \ref{fig1}(b) shows a representative SEM image of the fabricated SILs, while Figure \ref{fig1}(d) presents the spatial PL map of the defect emitter integrated within a SIL. Further details on the fabrication and characterization of SILs can be found in previous studies\cite{geng2023dephasing,geng2023optical,geng2024defect}.

We conducted the temperature-dependent polarization experiment using a custom-built confocal scanning microscope setup as shown in Figure \ref{fig2}. The GaN sample was mounted in a helium-flow cryostat, with the temperature controlled from 10 K to 300 K (room temperature). A 532 nm continuous wave laser was used to excite the sample. The PL light from the defect emitter was collected by an objective lens, passed through a confocal spatial filter, and directed to a spectrometer and an HBT interferometer for PL spectrum and $g^{(2)}(\tau)$ measurements. A half-wave plate and a polarizer were inserted into the collection path to analyze the polarization characteristics of the PL emission. More detailed information on the experimental setup can be found in previous studies\cite{geng2023dephasing,geng2023optical}.

\begin{figure}[htb]
		\includegraphics[width=1\columnwidth]{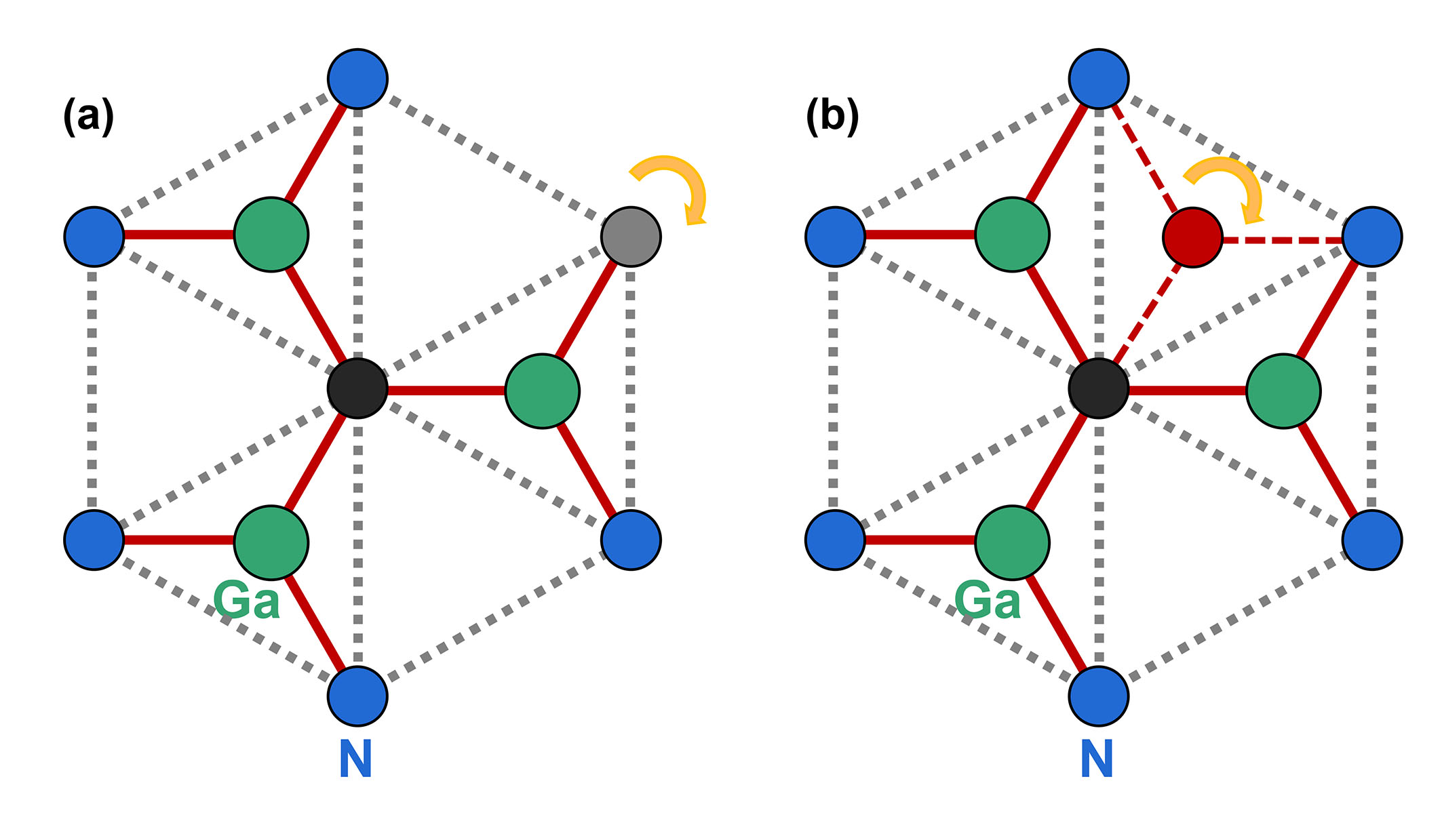}
        \centering
		\caption{ Top view of the GaN crystal. (a) Substitutional defect indicated by gray and black dots. (b) Interstitial defect represented by a red dot.}
		\label{fig4}
	\end{figure}

Figure \ref{fig3} presents the results of the temperature-dependent emission polarization of the GaN defect emitter. The experimental data (blue dots) are well fitted by the function $cos^{2}(\phi-\phi_{0})$ (red curves), where $\phi_{0}$ corresponds to the polarization direction of the PL emission. First, we observe that the light emitted from the defect emitter remains linearly polarized throughout the temperature range, confirming that the emitter follows the single optical dipole model. Second, while the emission remains linearly polarized, the polarization direction shifts at different temperatures. As shown in Figure \ref{fig3}, at 10 K, the polarization direction is at $139^{\circ}$, and as the temperature increases, it gradually shifts to $154^{\circ}$ at room temperature. This behavior is entirely different from that of the NV center in diamond, whose excited state is an orbital doublet, allowing the emission of two orthogonal polarizations ($E_{x}$ and $E_{y}$)\cite{fu2009observation}. Third, we observe that below 100 K, the polarization direction of the GaN defect emitter remains nearly unchanged as the temperature increases, while a more significant rotation occurs in the temperature range of 100-200 K. Previous studies, based on symmetry considerations, have suggested that the nature of the GaN defect emitter cannot be attributed to a single substitutional or interstitial impurity atom. Instead, it is most likely a defect complex composed of at least two or more impurity atoms or vacancies. Substitutional impurity atoms have bond lengths that differ from those of Ga and N atoms in the bulk crystal, leading to local lattice distortions\cite{matsubara2017first}. Figure \ref{fig4}(a) shows a top view of the GaN crystal, with the Ga atoms colored green and the N atoms blue. The black and gray circles represent substitutional impurities or vacancies. The presence of a defect can break the crystal symmetry and induce local distortion, which may cause the dipole moment to deviate from alignment with the exact crystallographic direction. These configurations may also be temperature-dependent. If interstitial impurity atoms as shown in Figure \ref{fig4}(b) (red dot), such as carbon, are present, they may undergo positional shifts as the temperature changes. These factors could contribute to the rotation of the PL polarization direction as a function of temperature.

Clearly, further experiments are required to gain a deeper understanding of the polarization properties of the defect emitter. Future experimental studies may include the following aspects. First, it is essential to investigate a sufficiently large number of defect emitters to gather statistical information. The direction of polarization, whether it rotates, the rotation angle, and the temperature range in which the rotation occurs should be correlated with the orientation of the GaN crystal and the center wavelength of the defect emitters. This approach can aid in further classifying GaN defect emitters, providing valuable insights into their underlying nature. Second, for the same defect emitter, the polarization characteristics of its PL can be investigated under repeated heating and cooling cycles. Specifically, we can explore whether the rotation of the polarization direction occurs consistently or randomly. This would be particularly useful for understanding the defect configuration, especially in distinguishing between interstitial and substitutional impurities.

In summary, we employed a custom-built confocal scanning microscope to study the temperature-dependent emission polarization characteristics of GaN defect emitters integrated with solid immersion lenses. Our results confirm that GaN defect emitters follow the single optical dipole model in the 10-300K temperature range, although the direction of emission polarization may change with temperature. Goals and a roadmap for future research are also outlined. We hope that this work will inspire and stimulate further experimental and theoretical investigations in the field of GaN defect emitters.

\begin{acknowledgments}
This work was supported by the Cornell Center for Materials Research through the NSF MRSEC program (DMR-1719875), the NSF RAISE:TAQS program (ECCS-1839196), and the NSF EAGER program (CMMI-2240267).
\end{acknowledgments}

\nocite{*}
\bibliography{aipsamp}% Produces the bibliography via BibTeX.

\begin{thebibliography}{19}%
\makeatletter
\providecommand \@ifxundefined [1]{%
 \@ifx{#1\undefined}
}%
\providecommand \@ifnum [1]{%
 \ifnum #1\expandafter \@firstoftwo
 \else \expandafter \@secondoftwo
 \fi
}%
\providecommand \@ifx [1]{%
 \ifx #1\expandafter \@firstoftwo
 \else \expandafter \@secondoftwo
 \fi
}%
\providecommand \natexlab [1]{#1}%
\providecommand \enquote  [1]{``#1''}%
\providecommand \bibnamefont  [1]{#1}%
\providecommand \bibfnamefont [1]{#1}%
\providecommand \citenamefont [1]{#1}%
\providecommand \href@noop [0]{\@secondoftwo}%
\providecommand \href [0]{\begingroup \@sanitize@url \@href}%
\providecommand \@href[1]{\@@startlink{#1}\@@href}%
\providecommand \@@href[1]{\endgroup#1\@@endlink}%
\providecommand \@sanitize@url [0]{\catcode `\\12\catcode `\$12\catcode `\&12\catcode `\#12\catcode `\^12\catcode `\_12\catcode `\%12\relax}%
\providecommand \@@startlink[1]{}%
\providecommand \@@endlink[0]{}%
\providecommand \url  [0]{\begingroup\@sanitize@url \@url }%
\providecommand \@url [1]{\endgroup\@href {#1}{\urlprefix }}%
\providecommand \urlprefix  [0]{URL }%
\providecommand \Eprint [0]{\href }%
\providecommand \doibase [0]{http://dx.doi.org/}%
\providecommand \selectlanguage [0]{\@gobble}%
\providecommand \bibinfo  [0]{\@secondoftwo}%
\providecommand \bibfield  [0]{\@secondoftwo}%
\providecommand \translation [1]{[#1]}%
\providecommand \BibitemOpen [0]{}%
\providecommand \bibitemStop [0]{}%
\providecommand \bibitemNoStop [0]{.\EOS\space}%
\providecommand \EOS [0]{\spacefactor3000\relax}%
\providecommand \BibitemShut  [1]{\csname bibitem#1\endcsname}%
\let\auto@bib@innerbib\@empty
%</preamble>
\bibitem [{\citenamefont {Berhane}\ \emph {et~al.}(2018)\citenamefont {Berhane}, \citenamefont {Jeong}, \citenamefont {Bradac}, \citenamefont {Walsh}, \citenamefont {Englund}, \citenamefont {Toth},\ and\ \citenamefont {Aharonovich}}]{berhane2018photophysics}%
  \BibitemOpen
  \bibfield  {author} {\bibinfo {author} {\bibfnamefont {A.~M.}\ \bibnamefont {Berhane}}, \bibinfo {author} {\bibfnamefont {K.-Y.}\ \bibnamefont {Jeong}}, \bibinfo {author} {\bibfnamefont {C.}~\bibnamefont {Bradac}}, \bibinfo {author} {\bibfnamefont {M.}~\bibnamefont {Walsh}}, \bibinfo {author} {\bibfnamefont {D.}~\bibnamefont {Englund}}, \bibinfo {author} {\bibfnamefont {M.}~\bibnamefont {Toth}}, \ and\ \bibinfo {author} {\bibfnamefont {I.}~\bibnamefont {Aharonovich}},\ }\bibfield  {title} {\enquote {\bibinfo {title} {Photophysics of gan single-photon emitters in the visible spectral range},}\ }\href@noop {} {\bibfield  {journal} {\bibinfo  {journal} {Physical Review B}\ }\textbf {\bibinfo {volume} {97}},\ \bibinfo {pages} {165202} (\bibinfo {year} {2018})}\BibitemShut {NoStop}%
\bibitem [{\citenamefont {Luo}\ \emph {et~al.}(2025)\citenamefont {Luo}, \citenamefont {Geng}, \citenamefont {Rana},\ and\ \citenamefont {Fuchs}}]{luo2025gan}%
  \BibitemOpen
  \bibfield  {author} {\bibinfo {author} {\bibfnamefont {J.}~\bibnamefont {Luo}}, \bibinfo {author} {\bibfnamefont {Y.}~\bibnamefont {Geng}}, \bibinfo {author} {\bibfnamefont {F.}~\bibnamefont {Rana}}, \ and\ \bibinfo {author} {\bibfnamefont {G.}~\bibnamefont {Fuchs}},\ }\bibfield  {title} {\enquote {\bibinfo {title} {Gan quantum emitters: basic physics, spin structure, and optically detected magnetic resonance (odmr)},}\ }in\ \href@noop {} {\emph {\bibinfo {booktitle} {Gallium Nitride Materials and Devices XX}}}\ (\bibinfo {organization} {SPIE},\ \bibinfo {year} {2025})\ p.\ \bibinfo {pages} {PC1336605}\BibitemShut {NoStop}%
\bibitem [{\citenamefont {Berhane}\ \emph {et~al.}(2016)\citenamefont {Berhane}, \citenamefont {Jeong}, \citenamefont {Bodrog}, \citenamefont {Fiedler}, \citenamefont {Schr{\"o}der}, \citenamefont {Trivi{\~n}o}, \citenamefont {Palacios}, \citenamefont {Gali}, \citenamefont {Toth}, \citenamefont {Englund} \emph {et~al.}}]{berhane2016bright}%
  \BibitemOpen
  \bibfield  {author} {\bibinfo {author} {\bibfnamefont {A.~M.}\ \bibnamefont {Berhane}}, \bibinfo {author} {\bibfnamefont {K.-Y.}\ \bibnamefont {Jeong}}, \bibinfo {author} {\bibfnamefont {Z.}~\bibnamefont {Bodrog}}, \bibinfo {author} {\bibfnamefont {S.}~\bibnamefont {Fiedler}}, \bibinfo {author} {\bibfnamefont {T.}~\bibnamefont {Schr{\"o}der}}, \bibinfo {author} {\bibfnamefont {N.~V.}\ \bibnamefont {Trivi{\~n}o}}, \bibinfo {author} {\bibfnamefont {T.}~\bibnamefont {Palacios}}, \bibinfo {author} {\bibfnamefont {A.}~\bibnamefont {Gali}}, \bibinfo {author} {\bibfnamefont {M.}~\bibnamefont {Toth}}, \bibinfo {author} {\bibfnamefont {D.}~\bibnamefont {Englund}},  \emph {et~al.},\ }\bibfield  {title} {\enquote {\bibinfo {title} {Bright room-temperature single photon emission from defects in gallium nitride},}\ }\href@noop {} {\bibfield  {journal} {\bibinfo  {journal} {arXiv preprint arXiv:1610.04692}\ } (\bibinfo {year} {2016})}\BibitemShut {NoStop}%
\bibitem [{\citenamefont {Luo}\ \emph {et~al.}(2024{\natexlab{a}})\citenamefont {Luo}, \citenamefont {Geng}, \citenamefont {Rana},\ and\ \citenamefont {Fuchs}}]{luo20241room}%
  \BibitemOpen
  \bibfield  {author} {\bibinfo {author} {\bibfnamefont {J.}~\bibnamefont {Luo}}, \bibinfo {author} {\bibfnamefont {Y.}~\bibnamefont {Geng}}, \bibinfo {author} {\bibfnamefont {F.}~\bibnamefont {Rana}}, \ and\ \bibinfo {author} {\bibfnamefont {G.}~\bibnamefont {Fuchs}},\ }\bibfield  {title} {\enquote {\bibinfo {title} {Room temperature optically detected magnetic resonance in single spins hosted in gan},}\ }\href@noop {} {\bibfield  {journal} {\bibinfo  {journal} {Bulletin of the American Physical Society}\ } (\bibinfo {year} {2024}{\natexlab{a}})}\BibitemShut {NoStop}%
\bibitem [{\citenamefont {Geng}(2024)}]{geng2024defect}%
  \BibitemOpen
  \bibfield  {author} {\bibinfo {author} {\bibfnamefont {Y.}~\bibnamefont {Geng}},\ }\emph {\bibinfo {title} {Defect Quantum Emitters in Gallium Nitride}},\ \href@noop {} {Ph.D. thesis},\ \bibinfo  {school} {Cornell University} (\bibinfo {year} {2024})\BibitemShut {NoStop}%
\bibitem [{\citenamefont {Luo}\ \emph {et~al.}(2024{\natexlab{b}})\citenamefont {Luo}, \citenamefont {Geng}, \citenamefont {Farhan}, \citenamefont {Fuchs} \emph {et~al.}}]{luo2024data}%
  \BibitemOpen
  \bibfield  {author} {\bibinfo {author} {\bibfnamefont {J.}~\bibnamefont {Luo}}, \bibinfo {author} {\bibfnamefont {Y.}~\bibnamefont {Geng}}, \bibinfo {author} {\bibfnamefont {R.}~\bibnamefont {Farhan}}, \bibinfo {author} {\bibfnamefont {G.}~\bibnamefont {Fuchs}},  \emph {et~al.},\ }\bibfield  {title} {\enquote {\bibinfo {title} {Data and scripts from: Room temperature optically detected magnetic resonance of single spins in gan},}\ }\href@noop {} {\bibfield  {journal} {\bibinfo  {journal} {Cornell eCommons}\ } (\bibinfo {year} {2024}{\natexlab{b}})}\BibitemShut {NoStop}%
\bibitem [{\citenamefont {Luo}\ \emph {et~al.}(2023)\citenamefont {Luo}, \citenamefont {Geng}, \citenamefont {Rana},\ and\ \citenamefont {Fuchs}}]{luo2023room}%
  \BibitemOpen
  \bibfield  {author} {\bibinfo {author} {\bibfnamefont {J.}~\bibnamefont {Luo}}, \bibinfo {author} {\bibfnamefont {Y.}~\bibnamefont {Geng}}, \bibinfo {author} {\bibfnamefont {F.}~\bibnamefont {Rana}}, \ and\ \bibinfo {author} {\bibfnamefont {G.~D.}\ \bibnamefont {Fuchs}},\ }\bibfield  {title} {\enquote {\bibinfo {title} {Room-temperature optically detected magnetic resonance of gan defect single-photon emitters},}\ }in\ \href@noop {} {\emph {\bibinfo {booktitle} {CLEO: Fundamental Science}}}\ (\bibinfo {organization} {Optica Publishing Group},\ \bibinfo {year} {2023})\ pp.\ \bibinfo {pages} {FM1E--5}\BibitemShut {NoStop}%
\bibitem [{\citenamefont {Geng}\ \emph {et~al.}(2022{\natexlab{a}})\citenamefont {Geng}, \citenamefont {Luo}, \citenamefont {van Deurzen}, \citenamefont {Jena}, \citenamefont {Fuchs}, \citenamefont {Rana} \emph {et~al.}}]{geng2022decoherence}%
  \BibitemOpen
  \bibfield  {author} {\bibinfo {author} {\bibfnamefont {Y.}~\bibnamefont {Geng}}, \bibinfo {author} {\bibfnamefont {J.}~\bibnamefont {Luo}}, \bibinfo {author} {\bibfnamefont {L.}~\bibnamefont {van Deurzen}}, \bibinfo {author} {\bibfnamefont {D.}~\bibnamefont {Jena}}, \bibinfo {author} {\bibfnamefont {G.~D.}\ \bibnamefont {Fuchs}}, \bibinfo {author} {\bibfnamefont {F.}~\bibnamefont {Rana}},  \emph {et~al.},\ }\bibfield  {title} {\enquote {\bibinfo {title} {Decoherence by optical phonons in gan defect single-photon emitters},}\ }\href@noop {} {\bibfield  {journal} {\bibinfo  {journal} {arXiv preprint arXiv:2206.12636}\ } (\bibinfo {year} {2022}{\natexlab{a}})}\BibitemShut {NoStop}%
\bibitem [{\citenamefont {Wasisto}\ \emph {et~al.}(2019)\citenamefont {Wasisto}, \citenamefont {Prades}, \citenamefont {G{\"u}link},\ and\ \citenamefont {Waag}}]{wasisto2019beyond}%
  \BibitemOpen
  \bibfield  {author} {\bibinfo {author} {\bibfnamefont {H.~S.}\ \bibnamefont {Wasisto}}, \bibinfo {author} {\bibfnamefont {J.~D.}\ \bibnamefont {Prades}}, \bibinfo {author} {\bibfnamefont {J.}~\bibnamefont {G{\"u}link}}, \ and\ \bibinfo {author} {\bibfnamefont {A.}~\bibnamefont {Waag}},\ }\bibfield  {title} {\enquote {\bibinfo {title} {Beyond solid-state lighting: Miniaturization, hybrid integration, and applications of gan nano-and micro-leds},}\ }\href@noop {} {\bibfield  {journal} {\bibinfo  {journal} {Applied Physics Reviews}\ }\textbf {\bibinfo {volume} {6}} (\bibinfo {year} {2019})}\BibitemShut {NoStop}%
\bibitem [{\citenamefont {Yi-Fei}\ \emph {et~al.}(2019)\citenamefont {Yi-Fei}, \citenamefont {Zhu-Ning}, \citenamefont {Yao-Guang},\ and\ \citenamefont {Fei}}]{yi2019topological}%
  \BibitemOpen
  \bibfield  {author} {\bibinfo {author} {\bibfnamefont {G.}~\bibnamefont {Yi-Fei}}, \bibinfo {author} {\bibfnamefont {W.}~\bibnamefont {Zhu-Ning}}, \bibinfo {author} {\bibfnamefont {M.}~\bibnamefont {Yao-Guang}}, \ and\ \bibinfo {author} {\bibfnamefont {G.}~\bibnamefont {Fei}},\ }\bibfield  {title} {\enquote {\bibinfo {title} {Topological surface plasmon. polaritons},}\ }\href@noop {} {\bibfield  {journal} {\bibinfo  {journal} {Acta Physica Sinica}\ }\textbf {\bibinfo {volume} {68}} (\bibinfo {year} {2019})}\BibitemShut {NoStop}%
\bibitem [{\citenamefont {Najda}\ \emph {et~al.}(2022)\citenamefont {Najda}, \citenamefont {Perlin}, \citenamefont {Suski}, \citenamefont {Marona}, \citenamefont {Leszczy{\'n}ski}, \citenamefont {Wisniewski}, \citenamefont {Stanczyk}, \citenamefont {Schiavon}, \citenamefont {Slight}, \citenamefont {Watson} \emph {et~al.}}]{najda2022gan}%
  \BibitemOpen
  \bibfield  {author} {\bibinfo {author} {\bibfnamefont {S.~P.}\ \bibnamefont {Najda}}, \bibinfo {author} {\bibfnamefont {P.}~\bibnamefont {Perlin}}, \bibinfo {author} {\bibfnamefont {T.}~\bibnamefont {Suski}}, \bibinfo {author} {\bibfnamefont {L.}~\bibnamefont {Marona}}, \bibinfo {author} {\bibfnamefont {M.}~\bibnamefont {Leszczy{\'n}ski}}, \bibinfo {author} {\bibfnamefont {P.}~\bibnamefont {Wisniewski}}, \bibinfo {author} {\bibfnamefont {S.}~\bibnamefont {Stanczyk}}, \bibinfo {author} {\bibfnamefont {D.}~\bibnamefont {Schiavon}}, \bibinfo {author} {\bibfnamefont {T.}~\bibnamefont {Slight}}, \bibinfo {author} {\bibfnamefont {M.~A.}\ \bibnamefont {Watson}},  \emph {et~al.},\ }\bibfield  {title} {\enquote {\bibinfo {title} {Gan laser diode technology for visible-light communications},}\ }\href@noop {} {\bibfield  {journal} {\bibinfo  {journal} {Electronics}\ }\textbf {\bibinfo {volume} {11}},\ \bibinfo {pages} {1430} (\bibinfo {year} {2022})}\BibitemShut {NoStop}%
\bibitem [{\citenamefont {Geng}\ \emph {et~al.}(2022{\natexlab{b}})\citenamefont {Geng}, \citenamefont {Luo}, \citenamefont {van Deurzen}, \citenamefont {Jena}, \citenamefont {Fuchs},\ and\ \citenamefont {Rana}}]{geng20221temperature}%
  \BibitemOpen
  \bibfield  {author} {\bibinfo {author} {\bibfnamefont {Y.}~\bibnamefont {Geng}}, \bibinfo {author} {\bibfnamefont {J.}~\bibnamefont {Luo}}, \bibinfo {author} {\bibfnamefont {L.}~\bibnamefont {van Deurzen}}, \bibinfo {author} {\bibfnamefont {D.}~\bibnamefont {Jena}}, \bibinfo {author} {\bibfnamefont {G.}~\bibnamefont {Fuchs}}, \ and\ \bibinfo {author} {\bibfnamefont {F.}~\bibnamefont {Rana}},\ }\bibfield  {title} {\enquote {\bibinfo {title} {Temperature dependence of spectral emission from gan defect quantum emitters},}\ }\href@noop {} {\bibfield  {journal} {\bibinfo  {journal} {Bulletin of the American Physical Society}\ }\textbf {\bibinfo {volume} {67}} (\bibinfo {year} {2022}{\natexlab{b}})}\BibitemShut {NoStop}%
\bibitem [{\citenamefont {Geng}\ and\ \citenamefont {Nomoto}(2023)}]{geng2023ultrafast}%
  \BibitemOpen
  \bibfield  {author} {\bibinfo {author} {\bibfnamefont {Y.}~\bibnamefont {Geng}}\ and\ \bibinfo {author} {\bibfnamefont {K.}~\bibnamefont {Nomoto}},\ }\bibfield  {title} {\enquote {\bibinfo {title} {Ultrafast spectral diffusion of gan defect single photon emitters},}\ }\href@noop {} {\bibfield  {journal} {\bibinfo  {journal} {Applied Physics Letters}\ }\textbf {\bibinfo {volume} {123}} (\bibinfo {year} {2023})}\BibitemShut {NoStop}%
\bibitem [{\citenamefont {Luo}\ \emph {et~al.}(2024{\natexlab{c}})\citenamefont {Luo}, \citenamefont {Geng}, \citenamefont {Rana},\ and\ \citenamefont {Fuchs}}]{luo2024room}%
  \BibitemOpen
  \bibfield  {author} {\bibinfo {author} {\bibfnamefont {J.}~\bibnamefont {Luo}}, \bibinfo {author} {\bibfnamefont {Y.}~\bibnamefont {Geng}}, \bibinfo {author} {\bibfnamefont {F.}~\bibnamefont {Rana}}, \ and\ \bibinfo {author} {\bibfnamefont {G.~D.}\ \bibnamefont {Fuchs}},\ }\bibfield  {title} {\enquote {\bibinfo {title} {Room temperature optically detected magnetic resonance of single spins in gan},}\ }\href@noop {} {\bibfield  {journal} {\bibinfo  {journal} {Nature Materials}\ }\textbf {\bibinfo {volume} {23}},\ \bibinfo {pages} {512--518} (\bibinfo {year} {2024}{\natexlab{c}})}\BibitemShut {NoStop}%
\bibitem [{\citenamefont {Geng}\ \emph {et~al.}(2022{\natexlab{c}})\citenamefont {Geng}, \citenamefont {Luo}, \citenamefont {van Deurzen}, \citenamefont {Jena}, \citenamefont {Fuchs}, \citenamefont {Rana} \emph {et~al.}}]{geng2022temperature}%
  \BibitemOpen
  \bibfield  {author} {\bibinfo {author} {\bibfnamefont {Y.}~\bibnamefont {Geng}}, \bibinfo {author} {\bibfnamefont {J.}~\bibnamefont {Luo}}, \bibinfo {author} {\bibfnamefont {L.}~\bibnamefont {van Deurzen}}, \bibinfo {author} {\bibfnamefont {D.}~\bibnamefont {Jena}}, \bibinfo {author} {\bibfnamefont {G.~D.}\ \bibnamefont {Fuchs}}, \bibinfo {author} {\bibfnamefont {F.}~\bibnamefont {Rana}},  \emph {et~al.},\ }\bibfield  {title} {\enquote {\bibinfo {title} {Temperature dependence of the emission spectrum of gan defect single-photon emitters},}\ }\href@noop {} {\bibfield  {journal} {\bibinfo  {journal} {arXiv e-prints}\ ,\ \bibinfo {pages} {arXiv--2206}} (\bibinfo {year} {2022}{\natexlab{c}})}\BibitemShut {NoStop}%
\bibitem [{\citenamefont {Geng}\ \emph {et~al.}(2023{\natexlab{a}})\citenamefont {Geng}, \citenamefont {Luo}, \citenamefont {van Deurzen}, \citenamefont {Xing}, \citenamefont {Jena}, \citenamefont {Fuchs},\ and\ \citenamefont {Rana}}]{geng2023dephasing}%
  \BibitemOpen
  \bibfield  {author} {\bibinfo {author} {\bibfnamefont {Y.}~\bibnamefont {Geng}}, \bibinfo {author} {\bibfnamefont {J.}~\bibnamefont {Luo}}, \bibinfo {author} {\bibfnamefont {L.}~\bibnamefont {van Deurzen}}, \bibinfo {author} {\bibfnamefont {H.}~\bibnamefont {Xing}}, \bibinfo {author} {\bibfnamefont {D.}~\bibnamefont {Jena}}, \bibinfo {author} {\bibfnamefont {G.~D.}\ \bibnamefont {Fuchs}}, \ and\ \bibinfo {author} {\bibfnamefont {F.}~\bibnamefont {Rana}},\ }\bibfield  {title} {\enquote {\bibinfo {title} {Dephasing by optical phonons in gan defect single-photon emitters},}\ }\href@noop {} {\bibfield  {journal} {\bibinfo  {journal} {Scientific Reports}\ }\textbf {\bibinfo {volume} {13}},\ \bibinfo {pages} {8678} (\bibinfo {year} {2023}{\natexlab{a}})}\BibitemShut {NoStop}%
\bibitem [{\citenamefont {Geng}\ \emph {et~al.}(2023{\natexlab{b}})\citenamefont {Geng}, \citenamefont {Jena}, \citenamefont {Fuchs}, \citenamefont {Zipfel},\ and\ \citenamefont {Rana}}]{geng2023optical}%
  \BibitemOpen
  \bibfield  {author} {\bibinfo {author} {\bibfnamefont {Y.}~\bibnamefont {Geng}}, \bibinfo {author} {\bibfnamefont {D.}~\bibnamefont {Jena}}, \bibinfo {author} {\bibfnamefont {G.~D.}\ \bibnamefont {Fuchs}}, \bibinfo {author} {\bibfnamefont {W.~R.}\ \bibnamefont {Zipfel}}, \ and\ \bibinfo {author} {\bibfnamefont {F.}~\bibnamefont {Rana}},\ }\bibfield  {title} {\enquote {\bibinfo {title} {Optical dipole structure and orientation of gan defect single-photon emitters},}\ }\href@noop {} {\bibfield  {journal} {\bibinfo  {journal} {ACS Photonics}\ }\textbf {\bibinfo {volume} {10}},\ \bibinfo {pages} {3723--3729} (\bibinfo {year} {2023}{\natexlab{b}})}\BibitemShut {NoStop}%
\bibitem [{\citenamefont {Fu}\ \emph {et~al.}(2009)\citenamefont {Fu}, \citenamefont {Santori}, \citenamefont {Barclay}, \citenamefont {Rogers}, \citenamefont {Manson},\ and\ \citenamefont {Beausoleil}}]{fu2009observation}%
  \BibitemOpen
  \bibfield  {author} {\bibinfo {author} {\bibfnamefont {K.-M.~C.}\ \bibnamefont {Fu}}, \bibinfo {author} {\bibfnamefont {C.}~\bibnamefont {Santori}}, \bibinfo {author} {\bibfnamefont {P.~E.}\ \bibnamefont {Barclay}}, \bibinfo {author} {\bibfnamefont {L.~J.}\ \bibnamefont {Rogers}}, \bibinfo {author} {\bibfnamefont {N.~B.}\ \bibnamefont {Manson}}, \ and\ \bibinfo {author} {\bibfnamefont {R.~G.}\ \bibnamefont {Beausoleil}},\ }\bibfield  {title} {\enquote {\bibinfo {title} {Observation of the dynamic jahn-teller effect in the excited states of nitrogen-vacancy centers in diamond},}\ }\href@noop {} {\bibfield  {journal} {\bibinfo  {journal} {Physical Review Letters}\ }\textbf {\bibinfo {volume} {103}},\ \bibinfo {pages} {256404} (\bibinfo {year} {2009})}\BibitemShut {NoStop}%
\bibitem [{\citenamefont {Matsubara}\ and\ \citenamefont {Bellotti}(2017)}]{matsubara2017first}%
  \BibitemOpen
  \bibfield  {author} {\bibinfo {author} {\bibfnamefont {M.}~\bibnamefont {Matsubara}}\ and\ \bibinfo {author} {\bibfnamefont {E.}~\bibnamefont {Bellotti}},\ }\bibfield  {title} {\enquote {\bibinfo {title} {A first-principles study of carbon-related energy levels in gan. ii. complexes formed by carbon and hydrogen, silicon or oxygen},}\ }\href@noop {} {\bibfield  {journal} {\bibinfo  {journal} {Journal of Applied Physics}\ }\textbf {\bibinfo {volume} {121}} (\bibinfo {year} {2017})}\BibitemShut {NoStop}%
\end{thebibliography}%

\end{document}